\documentclass[%
 reprint,
superscriptaddress,
 amsmath,amssymb,
 aps,
prb,
floatfix,
]{revtex4-2}
\usepackage{ulem} 
\usepackage[caption=false]{subfig}
\usepackage{graphicx}
\usepackage{dcolumn}
\usepackage{color}
\usepackage{bm}
\usepackage{hyperref}
\usepackage{tikz,xcolor,hyperref}
\definecolor{lime}{HTML}{A6CE39} 
\DeclareRobustCommand{\orcidicon}{
	\begin{tikzpicture}
	\draw[lime, fill=lime] (0,0) 
	circle [radius=0.16] 
	node[white] {{\fontfamily{qag}\selectfont \tiny ID}};
	\draw[white, fill=white] (-0.0625,0.095) 
	circle [radius=0.007];
	\end{tikzpicture}
	\hspace{-2mm}
} 
\foreach \x in {A, ..., Z}{\expandafter\xdef\csname orcid\x\endcsname{\noexpand\href{https://orcid.org/\csname orcidauthor\x\endcsname}
			{\noexpand\orcidicon}}
}

\begin{document}

\preprint{APS/123-QED}

\title{Impact of impurity scattering on odd-frequency 
spin-triplet pairing near the edge of the Kitaev Chain}

\author{\orcidA{Sparsh Mishra}}
\affiliation{%
 Department of Physics, Nagoya University, Nagoya 464-8602, Japan
}%
\author{Shun Tamura}%
\affiliation{%
  Department of Applied Physics, Nagoya University, Nagoya 464-8603, Japan
}%
\author{Akito Kobayashi}
\affiliation{%
Department of Physics, Nagoya University, Nagoya 464-8602, Japan }%

\author{Yukio Tanaka}
\affiliation{%
 Department of Applied Physics, Nagoya University, Nagoya 464-8603, Japan }%

\date{\today}

\begin{abstract}
We study a Kitaev chain model, which is the simplest model of topological superconductors hosting Majorana fermion, appearing as a zero-energy state at the edge. We analytically calculate the Green's function of the semi-infinite Kitaev chain with a delta-function-type impurity potential within 
the quasi-classical regime to obtain the spatial dependence of the induced odd-frequency pairing. It is found that if the position of the impurity is not far from the 
edge, the spatial profile of the local density of states (LDOS) and the odd-frequency spin-triplet $s$-wave 
pair amplitude is tunable as a function of the strength of the impurity 
potential. Moreover, the zero-energy LDOS and low-frequency odd-frequency pair amplitude are found to have the same spatial dependence. The spatial profile of the zero-energy LDOS is analyzed based on the wave function of Majorana fermions. 
\end{abstract}

\maketitle

\section{Introduction}

It is known that the symmetry of a Cooper pair plays a major role 
in determining the physical property of superconductivity. 
Conventionally, it is classified into spin-singlet even-parity or spin-triplet odd-parity. These pairings are referred to as even-frequency 
pairings where the pair amplitude does not have a sign change 
with the exchange of two time variables of the electrons 
that form a Cooper pair. 

However, an odd-frequency pairing, wherein the pair amplitude changes sign 
by this operation, has been proposed by 
Berezenskii \cite{berezinskii1974new} in the context of 
superfluid $^{3}$He. 
Odd-frequency pairings are different from the even-frequency ones because the fermions try to avoid each other in time
\cite{oddfreqrev1,oddfreqreview2}. 
Although the possible odd-frequency pairing has been studied in bulk 
strongly correlated systems \cite{scoddfreq1,scoddfreq2,scoddfreq3,Coleman2,Vojta,Fuseya,Hotta,Shigeta,Shigeta2,Solenov,Kusunose,Kusunose2}, 
it has been clarified that odd-frequency pairings are not easily realizable as a uniform superconducting state similar to even-frequency pairings 
\cite{Fominov2015}. 
Nonetheless, it has been established that 
odd-frequency pairings can be induced by the external symmetry breaking 
like exchange field \cite{oddfreqspinrot,Efetov2}, 
translational symmetry breaking \cite{oddfreqrev1}, 
and orbital hybridization \cite{multibandsc,multband2,multband3}, whereby the bulk and primary symmetry of the Cooper pair are the even-frequency one. 
It is known that in non-uniform superconducting systems, 
odd-frequency pairings are ubiquitously present 
and they become prominent in the presence of a zero-energy surface Andreev bound state (ZESABS) \cite{oddfreqrev1,nonuniform,oddfreqZESLDOS1,Tsintzis}. 
In diffusive normal metal / spin-triplet superconductor junctions, 
the anomalous proximity effect occurs owing to the emergence of the odd-frequency spin-triplet 
\textit{s}-wave pairing near the interface, which is robust against impurity scattering 
\cite{Proximityp,oddfreqinsight1,odd1}. 
This pairing also induces a paramagnetic Meissner response \cite{oddfreqinsight2,meissnertheory1,meissnertheory2,Suzukisanspaper,HigashitaniHTC} that is observed experimentally \cite{exp1,exp2,krieger2020proximityinduced}.  

Odd-frequency pairings have been attracting considerable interest, especially from the perspective of 
topological superconductors, wherein Majorana zero-energy states (MZESs) 
are generated as edge states \cite{oddfreqrev1}. 
The MZES is a certain type of ZESABS 
and it inevitably accompanies the odd-frequency spin-triplet $s$-wave pairing
near the edge  \cite{oddfreqrev1,PhysRevB.87.104513}. In the case of $p$-wave superconductivity, the odd-frequency spin-triplet $s$-wave pairing is generated in the presence of spatial non-uniformity \cite{oddfreqrev1,oddfreqorig,oddfreqZESLDOS1}.  
Recently, a more direct relation between the induced odd-frequency pairing 
and the bulk quantity has been derived by the spectral-bulk edge correspondence \cite{PhysRevB.99.184512,PhysRevB.100.174512}, which is an extended version of the 
bulk-edge correspondence \cite{Sato2011} derived for topological superconductors.  
The emergent MZES has special non-Abelian exchange statistics and 
is thought to provide new and powerful information processing methods \cite{wilczekss} and quantum computation schemes that are robust against impurity scattering \cite{Alicea_2012}. Since a minimal model that shows the emergence of MZES is the Kitaev chain\cite{Kitaev_2001}, it is important to clarify the stability of MZES on the application of external perturbation. 

The impact of impurity scattering on superconductivity has been a long-standing problem. One of the approaches to study the effect of impurity scattering in an \textit{s}-wave superconductor has been the use of Feynman diagram methods for various kinds of impurity potentials. {\color{black}{It has been shown \cite{PhysRevB.100.144511} for the \textit{s}-wave system that non-magnetic impurity potentials do not induce odd-frequency components as long as the spatial dependence of the $s$-wave pair potential is not influenced. On the other hand, odd-frequency components are induced by impurities possessing a quantum energy level.}} In a system with a single isolated magnetic impurity in a conventional $s$-wave superconductor, it was shown experimentally \cite{simon} that the odd-frequency spin-triplet component was enhanced near the impurity site because of rotational symmetry breaking. The impact of impurity scattering on ZESABS 
has been studied previously in normal metal/ unconventional superconductor junctions. 
When impurity scatterers are in the normal metal side. 
The proximity effect from ZESABS can only occur when the odd-frequency spin-triplet
\textit{s}-wave is generated at the interface \cite{Proximityp,oddfreqinsight1,odd1}. 
In the superconductor side,  
by considering uniform impurity scatterers with weak disorders 
based on the Eilenberger equation \cite{Eilenberger1968},  
it has been shown that the ZESABS in spin-triplet \textit{p}-wave superconductor 
junctions is robust against impurities because the odd-frequency spin-triplet $s$-wave pairing is generated \cite{PhysRevB.94.014504}. 
Beyond the weak disorders, there are several numerical calculations 
for \textit{p}-wave superconductors based on the tight binding model
\cite{takagi_tamura_tanaka_2020,impurityscattering2d}. However, studies on the impact of a strong impurity on the ZESABS remains limited \cite{Tinyukova2019}. 

In this work, we study the impact of impurity scattering on odd-frequency spin-triplet pairings near the edge of the Kitaev chain based on 
analytically obtained Green's functions. 
We find that if the position of the impurity is not far from the edge, the spatial 
profile of the local density of state (LDOS) and the odd-frequency spin-triplet $s$-wave pair amplitude can be 
tuned as a function of the strength of the impurity potential. \par
The remainder of this paper is organized as follows: in Sec. \ref{thmodel} we discuss the specific model and the scattering approach method used throughout this paper. Thereafter, we first show some simple non-uniform systems in Sec. \ref{nonuniformsec}. In Sec. \ref{results} we discuss the main results, namely, 
the LDOS of quasi-particles, the wave-function, and 
localisation length of the MZES\@. 
\section{Model and method}\label{thmodel}
\begin{figure}[htp]
\includegraphics[width=8.2 cm]{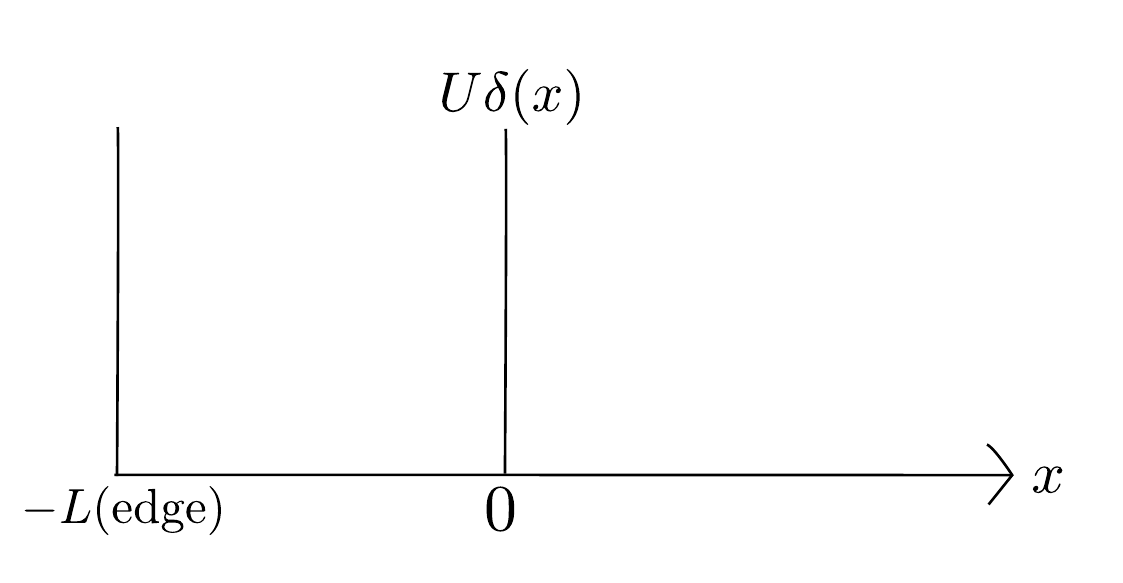}
\caption{\label{modelart} Schematic picture of semi-infinite system with impurity}
\end{figure}
In this section we introduce the model and method used herein.
\subsection{Model}
It is known that spin-triplet $p$-wave 
superconductors can host topological superconducting phase with 
ZESABS localized at the edge.  
To study the effect of a non-magnetic impurity on 
ZESABS, we consider a semi-infinite Kitaev chain, which is a model of  
fully polarized spin-triplet \textit{p}-wave superconductors. 
This is the simplest model of topological superconductivity
hosting Majorana fermions. 
As we are considering a fully polarized spin-triplet pairing; 
there is no spin degree of freedom in this model.
We consider this model in the continuum limit in the presence 
of a delta function impurity. 

Anisotropic superconductor systems can be described by solutions of the Bogoliubov de Gennes equation \cite{bdg,quasiclassicalbdg}. {\color{black}{We assume a mean-field Hamiltonian density $\hat{H}(x,x')$ for a $p$-wave spin-triplet superconductor with an impurity potential $\tilde{U}(x)$. The Hamiltonian $H$ is given by:
\begin{equation}
H = \int dx \int dx'C^\dagger (x) \hat{H}(x,x')C(x').
\end{equation}
Here, $C(x)$ is given by $C(x) = (c(x),c^\dagger (x))^T$ in the Nambu-spinor notation and $c(x)$ ($c(x)^\dagger$) is the annihilation (creation) operator for the Kitaev chain in the continuum limit.
We also have the following relations:}}
\begin{eqnarray}
    E \Psi(x)  &=& \int dx' \hat{H}(x,x')
    \Psi(x'), \label{BDgeq}
     \\
    \hat{H}(x,x') &=&  \begin{pmatrix}
     h(x,x') && \Delta(x,x') \\
     -\Delta^*(x,x') && -h(x,x')
    \end{pmatrix} , \label{nonlocalham} \\
    h(x,x') &=& \bigg(-\frac{\hslash^2}{2m}\frac{d^2}{dx^2}-\mu+ \tilde{U}(x) \bigg)\delta(x-x').
\end{eqnarray}
Here, $m$ is the mass of the electron, $\mu$ is the chemical potential, $\Delta(x,x')$ is the pair potential for the $p$-wave system, and $\tilde{U}(x)$ is the impurity potential defined as follows:
\begin{equation}
    \tilde{U}(x)  = 
    \begin{cases}
    \infty, & x < -L \\
    U \delta(x),& x \geq -L 
    \end{cases}
\end{equation}
where $U$ is the magnitude of the impurity potential.
{\color{black}{In Eq. (\ref{BDgeq}), $\Psi(x)$ is an eigenstate of $\hat{H}(x,x')$ given by Eq. (\ref{nonlocalham}). This wave-function is defined as: }}
\begin{equation}\label{wavefnnonlocal} 
    \Psi(x) \equiv \begin{pmatrix}
    u(x)\\v(x)
    \end{pmatrix} .
\end{equation}
A schematic of the model is shown in Fig. \ref{modelart}.

We impose the following boundary conditions on the wave-functions of the system:
\begin{align}
    \Psi(-L)=0, \label{bc1} \\
    \frac{d}{dx}\Psi(x)\bigl|_{0^-}^{0^+}=\frac{2m}{\hbar^2} U\Psi(0), \label{bc2} \\
    \Psi(0^+) = \Psi(0^-). \label{bc3}
\end{align}
To find analytic solutions, we solve the model within the quasi-classical limit $\mu \gg \Delta_0$. If we define the envelope wave-functions $\bar{\Psi}(\hat{k},x)$ as:
\begin{equation} \label{envelopfn}
   \bar{\Psi}(\hat{k},x) = 
     \begin{pmatrix}
    \bar{u}(\hat{k},x)\\\bar{v}(\hat{k},x)
    \end{pmatrix} 
     \equiv e^{-i k_f \hat{k} x}\begin{pmatrix}
     u(x) \\
     v(x)
    \end{pmatrix}, 
\end{equation}
Thereafter, we find that the Hamiltonian for the envelope functions in Eq. (\ref{envelopfn}) is as follows:

\begin{equation}
\label{QCham}
   \hat{H}_{QC}(\hat{k},x) = \bigg(-i \hbar v_f \hat{k} \frac{d}{d x} + \tilde{U}(x)\bigg)\sigma_z  + \Delta(\hat{k},x)\sigma_x  . 
\end{equation}
Here, $\hat{k} = k/k_f$ with the Fermi wave vector $k_f = \sqrt{2m\mu /\hbar^2}$, $v_f = \hbar k_f /m$, and $\sigma_i$ with $i = x,y,z$ is a Pauli matrix in Nambu space.
We assume that the pair potential is uniform. Then, it can be expressed as:
\begin{equation} \label{pwavedelta}
    \Delta(\hat{k},x) = 
   \Delta_0 \frac{k}{\sqrt{k^2}}.
\end{equation}
For simplicity we assume $\Im \Delta_0 = 0$ and $\Delta_0 >0$. 
More details about the derivation of Eq. (\ref{QCham}) are given in the Appendix \ref{Ham}. 
Finding the eigenvectors of $\hat{H}_{QC}$ and then using Eq. (\ref{envelopfn}) gives us the wave-functions of $\hat{H}(x,x')$ within the quasi-classical approximation. 

\subsection{Method}
We calculate a retarded Green's function for $\hat{H}(x,x')$ using a scattering approach \cite{McMillan} within the quasi-classical approximation. From the retarded Green's function, we can extract information about LDOS and the pair amplitude, which is used to analyze the symmetry of the Cooper pair. The extensive details of the method can be found in Appendices  \ref{wavefunctions},  \ref{scatteringstates}, \ref{scatteringcoef}, and \ref{greensfn}.

In the particle-hole space, the retarded Green's function for the system is given by a  $2 \times 2$ matrix:
\begin{eqnarray}
  \bm{G}^{r}(x,x',E) = \begin{pmatrix}
  G_{ee}^r & G_{eh}^r\\
    G_{he}^r & G_{hh}^r
  \end{pmatrix}.
\end{eqnarray}
The 11 component of the retarded Green's function gives us the LDOS\@, which is related to the LDOS through the following expression: 
$\rho(x,E) = -\frac{1}{\pi} \Im[G_{ee}^r(x,x,E)]$.
The 12 component of the Green's function in Nambu space is called the pair amplitude. 
As we are considering the spin-triplet superconductor without any 
external perturbation breaking spin-rotational symmetry, 
only spin-triplet pairing is 
allowed. Fermi-Dirac statistics dictates that 
an odd (even)-frequency pairing should have even (odd)-parity. 
Thus, the odd and even frequency components of the pair amplitudes are given by: 
\begin{equation}\label{oddfreq1}
    G_{odd}^{r}(x,x',E)\equiv \frac{G_{eh}^r(x,x',E) +G_{eh}^r(x',x,E)}{2},
\end{equation}
\begin{equation}\label{oddfreq2}
    G_{even}^{r}(x,x',E)\equiv \frac{G_{eh}^r(x,x',E) -G_{eh}^r(x',x,E)}{2} . 
\end{equation}
In principle, both even- and odd-frequency components exist, but for $x=x'$, 
the even frequency component vanishes, as seen in  Eq. (\ref{oddfreq2}).  
Therefore,  we get 
\begin{equation} \label{oddfreqeq1}
    G_{odd}^{r}(x,x,E) = G_{eh}^r(x,x,E) . 
\end{equation}

If we make the analytic continuation $E+ i\delta \rightarrow i\omega_n$ for the retarded Green's function (where $\delta$ is a positive infinitesimal and $\omega_n$ is the Matsubara frequency), 
$G_{even(odd)}(x,x',i\omega_{n})$ satisfy following equations: 
\begin{equation}
    G_{even}(x,x',-i\omega_{n}) = G_{even}(x,x',i\omega_{n}) , 
\end{equation}
\begin{equation}\label{oddfreqwhenxx'}
    G_{odd}(x,x',-i\omega_{n}) = - G_{odd}(x,x',i\omega_{n}) . 
\end{equation}
Eq. (\ref{oddfreqwhenxx'}), for $x=x'$ is the odd frequency spin-triplet $s$-wave pair amplitude. 

\section{Simple non-uniform systems} \label{nonuniformsec}
Before presenting the main results of this study, it is instructive to consider 
some simple non-uniform systems. We will examine the $p$-wave semi-infinite superconductor and the infinite $p$-wave superconductor with a single impurity.
\subsection{Semi-infinite geometry}
\begin{figure*}
\centering
\includegraphics[scale = 1]{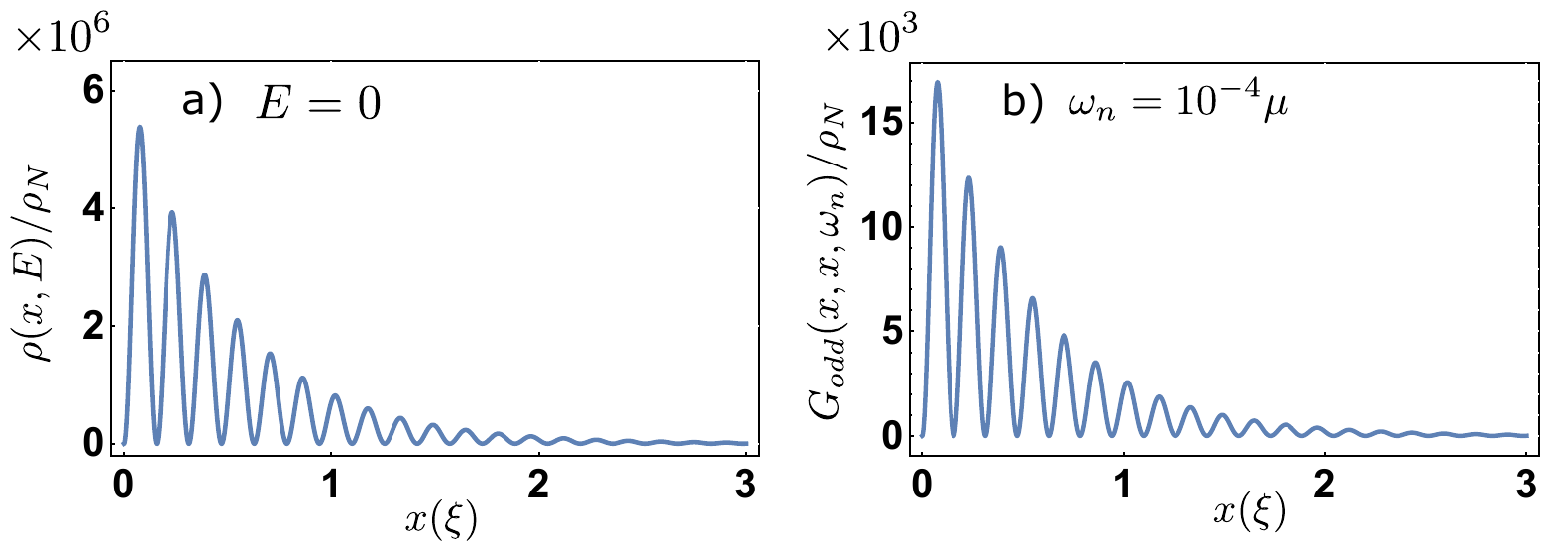}

\caption{\label{oddfreqldossemiinfinite} Spatial dependence of a) normalized 
LDOS at zero energy, b) normalized odd-frequency component of the pair amplitude for $x=x'$ for the semi-infinite Kitaev chain. We choose $\Delta_0/\mu=0.1$ and infinitesimal $\delta = 10^{-7}\mu$.}
\end{figure*}

The Green's function of the semi-infinite spin-triplet $p$-wave superconductor (superconductor present for $x>0$ with a boundary at $x=0$) within the quasi-classical approximation was calculated analytically 
\cite{PhysRevB.99.184512}. We  summarize some of the relevant results. The LDOS, the 12 component of the retarded Green's function, and its odd-frequency component for $x=x'$ are given by:
\begin{widetext}
\begin{eqnarray}
\rho(x,E) &=& \frac{-1}{\pi}\Im \bigg [\frac{m}{i k_f \hbar^2} 
 \bigg \{ \frac{E}{\Omega} - e^{2 i\gamma x}
[ \frac{E}{\Omega} \cos(2k_{f}x) + i \sin(2k_{F}x)]
-e^{2 i\gamma x}\frac{2 \Delta_0^{2}}{\Omega E} \sin^{2}(k_{f}x)
\bigg \} \bigg] ,\label{ldossemi1}\\
G_{eh}^r(x,x',E) &=& \frac{m}{i k_f \hbar^2} \bigg \{ \frac{\Delta_0}{\Omega} i \sin(k_f(x-x'))[e^{i \gamma |x-x'|} - e^{i \gamma (x+x')} ]  - 2 \frac{\Delta_0}{E} e^{i \gamma(x+x')}\sin(k_f x) \sin(k_f x') \bigg\} ,  \\
G_{odd}^r(x,x,E) &=& \frac{ 2 i m}{ k_f \hbar^2} \bigg \{ \frac{\Delta_0}{E} e^{i \gamma(2x)}\sin^2(k_f x) \bigg \}.
\end{eqnarray}
\end{widetext}
Here, $E$ denotes $E+i\delta$, $\Omega(E) = \sqrt{E^2 - \Delta_0^2}$, and $\gamma(E) = \frac{k_f \Omega(E)}{2\mu}$. 
If we write the corresponding LDOS for a semi-infinite spin-singlet $s$-wave superconductor, denoted as $\rho_s(x,E)$, we obtain: 
\begin{multline}
\rho_s(x,E) = \frac{-1}{\pi}\Im \bigg [\frac{m}{i k_f \hbar^2} 
 \bigg \{ \frac{E}{\Omega} - e^{ 2 i\gamma x}
[ \frac{E}{\Omega} \cos(2k_{f}x) \\+ i \sin(2k_{f}x)]
\bigg \} \bigg] .  \label{ldossemis1}
\end{multline}
Equation (\ref{ldossemis1}) does not contain the $1/E$ divergent term, which is from ZESABS\@, that is present in Eq. (\ref{ldossemi1}).
For the limiting case of $E = 0 + i\delta$, Eq. (\ref{ldossemi1}) can be approximated as:
\begin{equation}\label{lowenergysemi}
    \rho(x,0 + i\delta) = \frac{2 m \Delta_0 }{\delta k_f \hbar^2 \pi} e^{-2x/\xi}\sin^2(k_f x) . 
\end{equation} 
It is evident  that ZESABS is localized at the edge with the 
localization length $\xi$ (superconducting coherence length) given by: $\xi = \hbar v_f/ \Delta_0$, where $v_f$ is the Fermi velocity $v_f = \hbar k_f/m = (2/\hbar k_f) \mu$. 
The height of the LDOS 
then depends on the infinitesimal $\delta$.
The behavior of the odd frequency component is more apparent through analytic continuation $E+i\delta \rightarrow i\omega_n$, which makes it an odd function of the Matsubara frequency. If we choose $\omega_{n}$ as infinitesimal $\epsilon$, then $G_{odd}(x,x,\epsilon)$ 
becomes: 
\begin{equation}
G_{odd}(x,x,\epsilon)
= 
\frac{ 2 m \Delta_0}{ k_f \hbar^2 \epsilon} e^{-2x/\xi}\sin^2(k_f x) . 
\end{equation}
The spatial dependence of MZES is equivalent to the
odd-frequency spin-triplet $s$-wave pair amplitude generated at the edge 
\cite{PhysRevB.87.104513,takagi_tamura_tanaka_2020,oddfreqorig,oddfreqrev1}.
These features can be seen in Figs. \ref{oddfreqldossemiinfinite}(a) and \ref{oddfreqldossemiinfinite}(b). 
The $x$-axis is the position $x$ in units of the $\xi$, while the $y$-axis is normalized with respect to the normal metal density of states at zero energy $\rho_N$ ($\rho_N \equiv \frac{1}{2\pi^2}\frac{2m}{\hbar^2 k_f}$). The values of the pair potential and infinitesimal $\delta$ are taken as 
$\Delta_0 = 0.1\mu$ and $\delta = 10^{-7} \mu$, respectively.

\subsection{Infinite geometry with single impurity}
We now move on to the case where a single impurity is located at 
$x=0$ in a uniform Kitaev chain. 
Here, the impurity is modelled by a delta function $U\delta(x)$. 
We obtain the Green's function in Nambu space 
using a similar scattering technique. 
The LDOS and the odd-frequency component of the pair amplitude are given as follows: ($\Tilde{Z} = Z/k_f$ and $Z = \frac{2m}{\hslash^2}U$)
\begin{widetext}
\begin{multline}\label{ldosinfinitekitaev}
\rho(x,E) = \frac{-1}{\pi} \Im \bigg [ \frac{m}{ i\hbar^2 k_f } \bigg( \frac{E}{\Omega} - \frac{e^{2i \gamma|x| }E \sqrt{1 - \sigma_N (\tilde {Z})}}{  E^2  -  \Delta_0^2 \sigma_N(\tilde{Z})} \bigg \{ \sqrt{1 - \sigma_N (\tilde {Z})} \bigg[ \frac{2\Delta_0^2}{\Omega}\sin^2(k_f |x|) + \frac{E^2}{\Omega}\cos(2k_f |x|) + i E\sin(2k_f |x|) \bigg]  \\ +i\sqrt{\sigma_N (\tilde {Z})}\bigg[ E \cos(2k_f |x|) -i \Omega \sin(2 k_f |x|)\bigg] \bigg \}\bigg) \bigg ] . 
\end{multline}
\begin{eqnarray}
G_{odd}^r (x, x, E) = \frac {2 m i} {\hbar^2 k_f}e^{2 i\gamma | x |}\sin (k_f x) \frac {E \Delta_0 \sqrt{1 -\sigma_N(\tilde{Z})}} {E^2 - \Delta_0^2 {\sigma} _N (\tilde {Z})}\bigg\{\sqrt{\sigma_N (\tilde {Z})} \cos (k_f x) + \sqrt{1 - \sigma_N (\tilde {Z})}\sin (k_f | x |)\bigg\}. 
\end{eqnarray}
\end{widetext}
Where $\sigma_N(\tilde{Z})$ is the transparency of the normal metal junction given by:
\begin{equation}\label{boundstateeq2}
    \sigma_N(\tilde{Z}) \equiv \frac{4}{4+ \tilde{Z}^2}  . 
\end{equation}
Firstly, taking the limit $\tilde{Z}\rightarrow \infty$ and then $E \rightarrow 0$, we obtain:
\begin{equation}
  \rho(x,0+i\delta) = \frac{2m}{ \pi \hbar^2 k_f } \frac{ \Delta_0 }{\delta} e^{-2|x|/\xi}\sin^2(k_f |x|)  . 
\end{equation}
which is equivalent to Eq. (\ref{lowenergysemi}) for $x > 0$. For the opposite case i.e.\ taking the limit $E \rightarrow 0$ with finite value of $\tilde{Z}$, one obtains the zero energy LDOS $\rho(x,0+i\delta)=0$.

\begin{figure*} 
\centering
  \includegraphics[scale = 1]{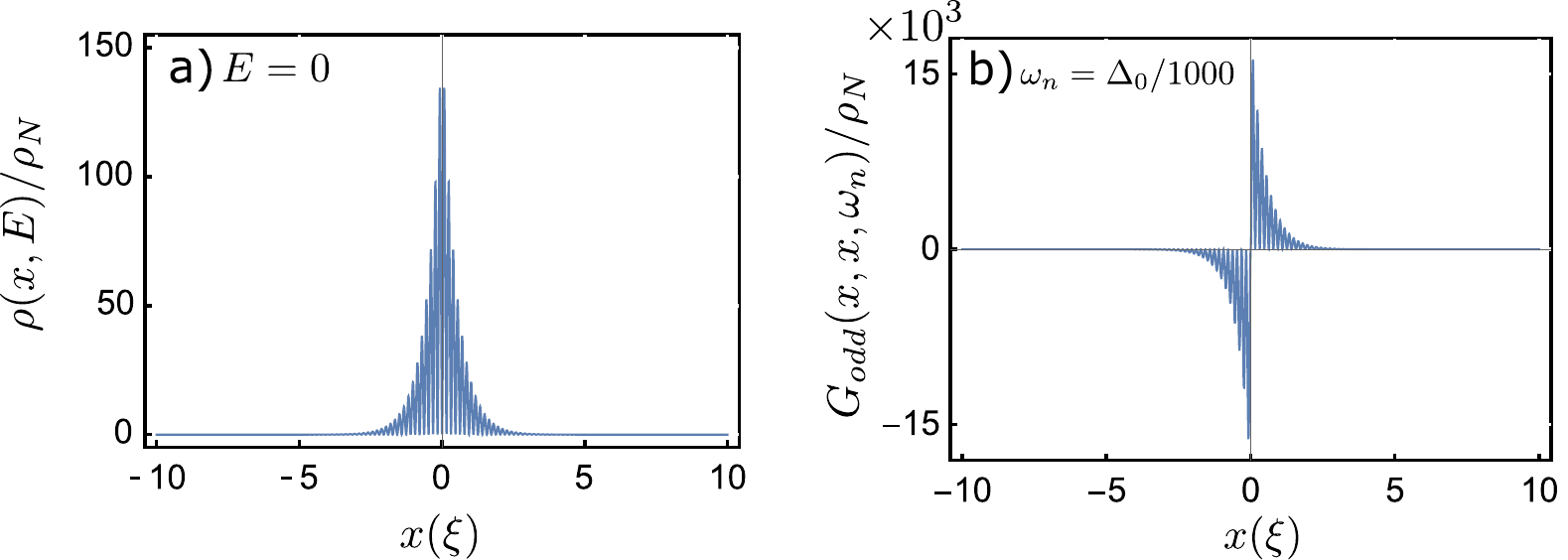}
\caption{ \label{LDOSoddfreqhomogeneous} Spatial dependence of a) Normalized 
LDOS at zero energy and  b) odd-frequency component of the pair amplitude 
for $x=x'$ at $\omega_n = \Delta_0/1000$ of the infinite Kitaev chain with a single impurity at $x=0$. $\Tilde{Z}=10^4$, $\Delta_0 = 0.1 \mu$
and infinitesimal $\delta = 10^{-7}\mu$ were used in the representative figures. }
\end{figure*}

Representative graphs of the zero-energy LDOS and the analytically continued odd-frequency component for $x=x'$ are shown in Figs. \ref{LDOSoddfreqhomogeneous}(a) and (b). 
The zero-energy LDOS (odd-frequency component) is symmetric (antisymmetric) about $x=0$.

Notably, $G_{odd}(x,x,E)$ has a sign change at $x=0$. 
For $\tilde{Z} \rightarrow \infty$ (or equivalently $\sigma_N(\tilde{Z}) \rightarrow 0$), the spatial dependence of $G_{odd}^r(x,x,E)$
is reduced to that for the semi-infinite Kitaev chain where the edge is located 
at $x=0$.
\subsubsection{Bound state for infinite system}\label{boundstatesection}

\begin{figure*} 
\centering
  \includegraphics{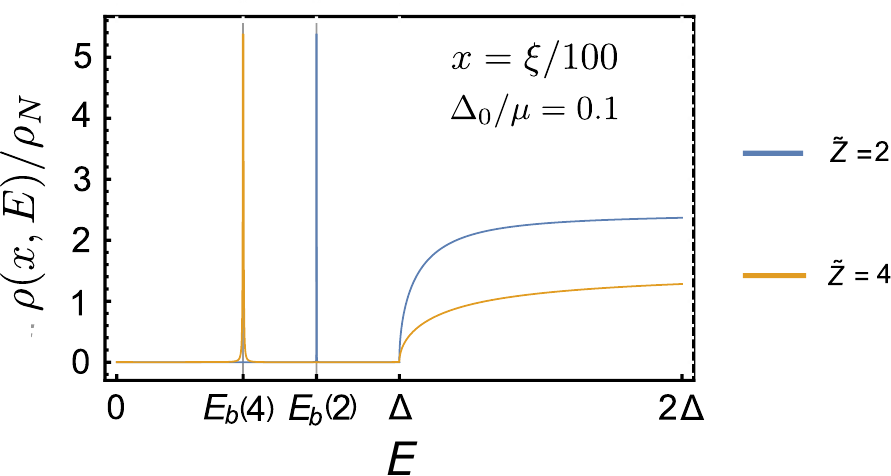}%
\caption{ \label{boundstate}LDOS at $x = \xi/100$ for infinite continuum Kitaev chain with a single impurity at $x=0$ for $\tilde{Z} = 2$ and $\tilde{Z} = 4$ with $\Delta_0=0.1 \mu$. 
}
\end{figure*}
For finite values of $\Tilde{Z}$, bound states exist at energies smaller than the gap $\Delta_0$. The infinite Kitaev chain model with an impurity at $x=0$ can be considered as a $p$-wave superconductor junction with non-zero transparency. We can then use the bound state expression obtained in the context of a $d$-wave- insulator $d$-wave junction\cite{dwave1,dwave2}:
\begin{equation} \label{boundstateeq1}
    E_b(\Tilde{Z}) = \Delta_0 \sqrt{\sigma_N(\Tilde{Z}}) . 
\end{equation}
The denominator of the second term in Eq. (\ref{ldosinfinitekitaev}) provides us with this bound state energy condition (see also Appendix \ref{scatteringcoef}). Figure \ref{boundstate} shows the LDOS, and the peaks that occur exactly at the expected bound state energies from Eq. (\ref{boundstateeq1}) and Eq. (\ref{boundstateeq2}).

\section{Semi-infinite geometry with impurity}\label{results}
To discuss the results of this study, we consider a system that is a combination of the two above-mentioned cases, i.e., a semi-infinite $p$-wave superconductor with an impurity. The Hamiltonian is given by Eq. (\ref{nonlocalham}), and the schematic is shown in Fig. \ref{modelart}. We focus on the topological regime wherein $\mu \gg \Delta_0$. 
\subsection{Local Density of States}\label{ldossection}
\begin{figure*}
\centering
  \includegraphics{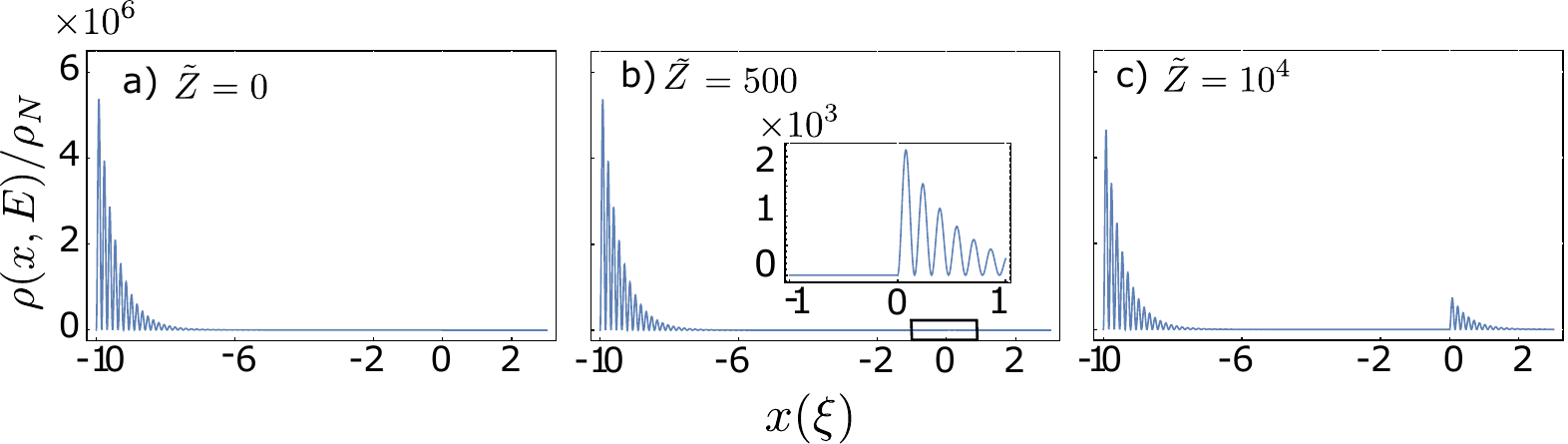}%
\caption{ \label{ldoslarge} Normalized LDOS for $E = 0+i\delta$, $L = 10 \xi$ with several values of $\tilde{Z}$ a) $\tilde{Z}=0$, b) $\tilde{Z}=500$ and c) $\tilde{Z}=10^4$. $\Delta_0 = 0.1\mu$. Positive infinitesimal $\delta = 10^{-7}\mu$.
Note that the $y$-axis is given in units of $10^{6}$ (semi-infinite system with impurity).}
\end{figure*}

\begin{figure*}
\centering
  \includegraphics{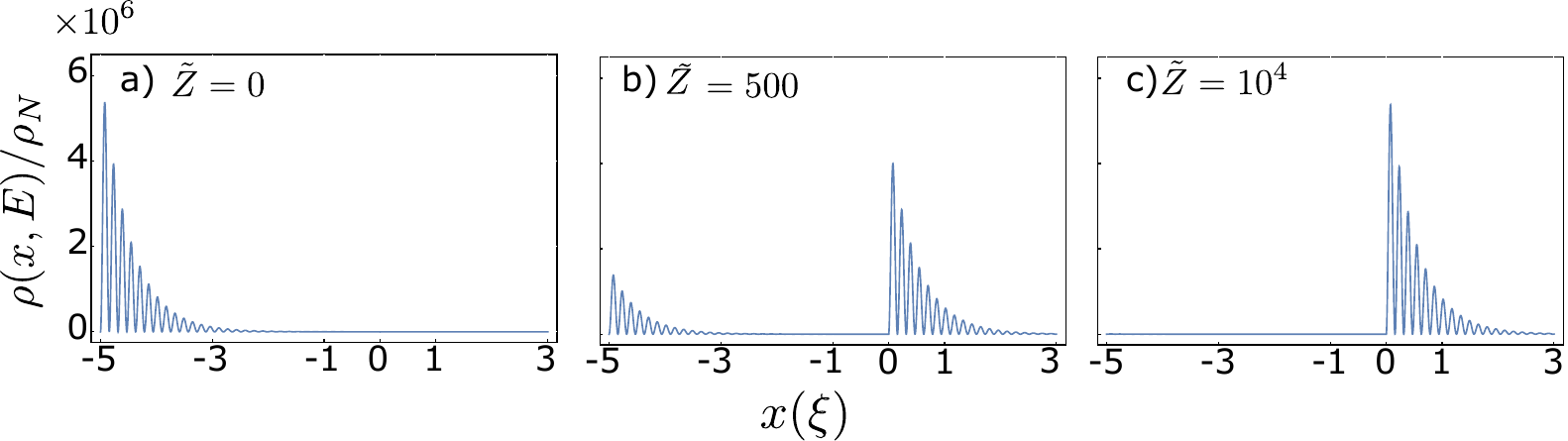}%
\caption{ \label{ldossmall}Normalized LDOS for $E = 0+i\delta$, $L = 5 \xi$ with several values of $\tilde{Z}$ a) $\tilde{Z}=0$, b) $\tilde{Z}=500$ and c) $\tilde{Z}=10^4$. $\Delta_0 = 0.1\mu$. Positive infinitesimal $\delta = 10^{-7}\mu$.
Note that the $y$-axis is given in units of $10^{6}$ (semi-infinite system with impurity).}
\end{figure*}

Firstly, we investigate how the energy spectrum of the system is altered because of the presence of an impurity. We start by finding the graph of the zero-energy LDOS. Here, the infinitesimal $\delta$ is chosen to be $10^{-7}\mu$. The plots are shown in Figs. \ref{ldoslarge} and \ref{ldossmall} wherein we have shown the normalized LDOS (normalized with respect to $\rho_N$, the density of states in normal metal) vs position $x$ in units of the superconducting coherence length. $\Delta_0$ is chosen to be $0.1\mu$. We define $L$ as the distance between the edge and impurity. We plotted for $L =10\xi $ and $L =5\xi $ in Figs. \ref{ldoslarge} and \ref{ldossmall}, respectively, for increasing values of impurity strengths $\Tilde{Z} = Z/k_f$ ($Z = \frac{2m}{\hbar^2}U$). Within the range of $\Tilde{Z}$ shown in the graphs, we see that as the impurity strength increases, so does the density of states at the right side of the impurity. We can understand this behavior by constructing the zero-energy state wave-function because the corresponding probability density must be qualitatively the same as the zero-energy LDOS. 

\subsection{Zero Energy state wave-function}\label{ZESsection}
\begin{figure*} 
\centering
  \includegraphics{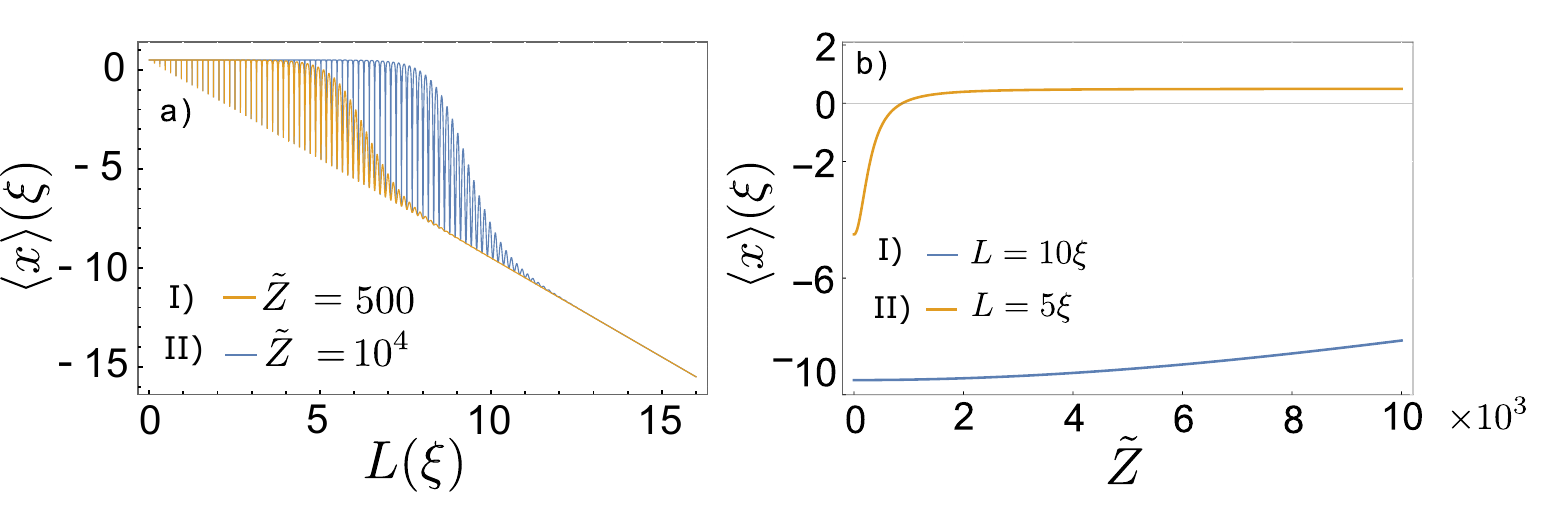}%
\caption{\label{mean} a) Mean position of wave function as a function of $L$ for I) $\tilde{Z} = 500$ and II) $\tilde{Z} = 10^4$. b) Mean position of wave function as a function of $\tilde{Z}$ for I) $L = 10\xi$ and II) $L = 5\xi$. Pair potential $\Delta_0=0.1\mu$. (Semi-infinite system with impurity)
}. 
\end{figure*}

We will now determine the zero-energy state wave function $\Psi_0(x)$ for the semi-infinite Kitaev system with the impurity. We take the wave function as a superposition of the $E=0$ eigenstates of the Hamiltonian given in Eq. (\ref{nonlocalham}).
We want solutions that decay at infinity and satisfy the boundary conditions in Eqs. (\ref{bc1}), (\ref{bc2}), and (\ref{bc3}). After performing the procedure, we obtain the following result:   
\begin{widetext}
\begin{eqnarray} \label{probzeroenergy}
   \Psi_0(x) = C \times
e^{- x/\xi} e^{-i k_f L}\bigg(\sin(k_f(x+L)) + \Theta(x) \tilde{Z} \sin(k_f L) \sin(k_f x)\bigg) \begin{pmatrix}
1 \\ -i
\end{pmatrix} ,
\end{eqnarray} 
\begin{equation} \label{constant}
     C = \bigg\{ \frac{\xi}{2 }\bigl[e^{2L/\xi}+ \tilde{Z}^2 \sin^2(k_f L) + \tilde{Z}\sin(2 k_f L) \bigl] \bigg\}^{-1/2}, 
\end{equation}
\end{widetext}
with the Heaviside step function $\Theta(x)$. We use the relation $1 \ll k_f \xi$, which is valid in the quasi-classical limit, which simplifies the expression considerably. 
From Eq. (\ref{probzeroenergy}), when $k_f L = n\pi$ with some integer $n$, the zero-energy state is not affected by the impurity. Subsequently, we consider $k_f L \neq n\pi$.
For $e^{L/\xi}/|\sin(k_f L)| \ll \Tilde{Z}$ (denoted as $\tilde{Z}\rightarrow \infty $), the probability density $|\Psi_0|^2_{\tilde{Z}\rightarrow \infty }(x)$ can be written as:
\begin{multline}\label{limz>>1}
  |\Psi_0|^2_{\tilde{Z}\rightarrow \infty }(x) \;=\; \bigg(\frac{4}{\xi}\bigg) e^{-2x/\xi}\sin^2(k_f x)\Theta(x) . 
\end{multline}
In the case of $\tilde{Z} = 0$, $|\Psi_0|^2_{\tilde{Z} = 0}(x)$ can straightforwardly be found from Eq. (\ref{probzeroenergy}):
\begin{multline}\label{limz<<1}
    |\Psi_0|^2_{\tilde{Z} = 0}(x) = \bigg(\frac{4}{\xi}\bigg)
     e^{-2(x+L)/\xi}\sin^2(k_f(x+L))  . 
\end{multline}
In Figs. \ref{ldoslarge} and \ref{ldossmall}, we show the zero-energy LDOS for $L = 10\xi$ and $L = 5\xi$, respectively. In these cases, $e^{L/\xi}/|\sin(k_f L)|$ is given by $e^{L/\xi}/|\sin(k_f L)| \sim 3 \times 10^4$ for $L = 10\xi$ and $\sim 3 \times 10^2$ for $L = 5\xi$. Then, $e^{L/\xi}/|\sin(k_f L)| \gg \tilde{Z}$ is satisfied for Figs. \ref{ldoslarge}(a), (b) and \ref{ldossmall} (a) and they can be explained by Eq. (\ref{limz<<1}). In addition, $e^{L/\xi}/|\sin(k_f L)| \ll \tilde{Z} $ is satisfied for Fig. \ref{ldossmall}(c) and it is explained by Eq. (\ref{limz<<1}). Figures \ref{ldoslarge}(c) and \ref{ldossmall}(b) are in the intermediate regime. 
From the obtained Eqs. (\ref{probzeroenergy}) and (\ref{constant}), we can see that on increasing the impurity strength, the zero-energy state becomes delocalized between the impurity site and the edge. Information on this delocalization can be extracted by evaluating the average position of the wave function, which is given by:
\begin{widetext}
\begin{eqnarray}
    \langle x \rangle &=& \frac{\xi}{2} + \frac{\xi }{2}\bigg\{ \frac{\frac{-2L}{\xi}e^{2L/\xi} + 4 \tilde{Z}\sin^2(k_f L)(\frac{1}{\xi k_f})^3}{e^{2L/\xi}+ \tilde{Z}^2 \sin^2(k_f L) + \tilde{Z}\sin(2 k_f L)} \bigg\} . 
\end{eqnarray}
\end{widetext}
where $\langle x \rangle$ is the mean position of the ZES\@. The asymptotic value of $\langle x \rangle$ for large $\Tilde{Z}$ ($e^{L/\xi}/|\sin(k_f L)| \ll \Tilde{Z}$) was found to be $\xi/2$. Figure \ref{mean}(a) gives the average position for $\tilde{Z} = 500$ and $\tilde{Z} = 10^4$ as a function of $L$\@. Figure \ref{mean}(b) shows the average position for $L = 10\xi$ and $L = 5\xi$ as a function of $\tilde{Z}$.

A peculiar feature of the ZES wave function and the LDOS in Figs. \ref{ldoslarge} and \ref{ldossmall} is that it is not symmetric locally around the impurity. This is in contrast to the LDOS of an isolated impurity in Fig. \ref{LDOSoddfreqhomogeneous}(a), which indicates a destructive interference between the waves scattered from the edge and those scattered from the impurity. 

\subsection{Odd-frequency component}\label{oddfreqsection}
\begin{figure*} 
\centering
  \includegraphics{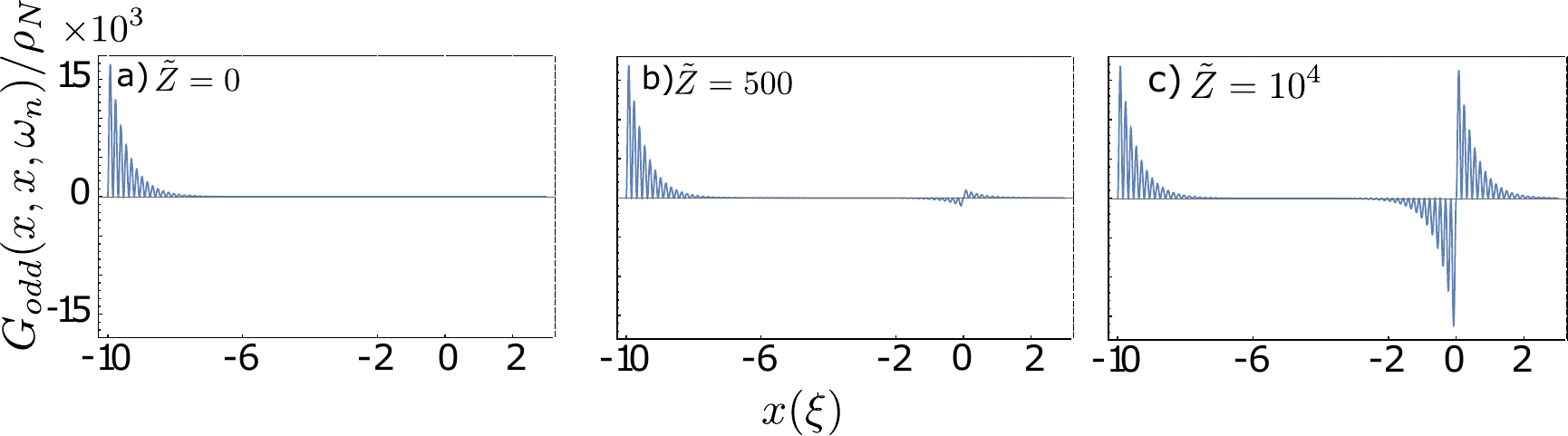}
\caption{ \label{oddfreqlarge} Normalized odd-frequency component of anomalous Green's function for $L = 10 \xi$ with several values of $\tilde{Z}$ a) $\tilde{Z}=0$, b) $\tilde{Z}=500$, and c) $\tilde{Z}=10^4$. The other parameters are $\Delta_0 = 0.1\mu$ and $\omega_n =\Delta_0/1000$.
Note that the $y$-axis is given in units of $10^{3}$ (semi-infinite system with impurity). }
\end{figure*}

\begin{figure*} 
\centering
  \includegraphics[scale = 1]{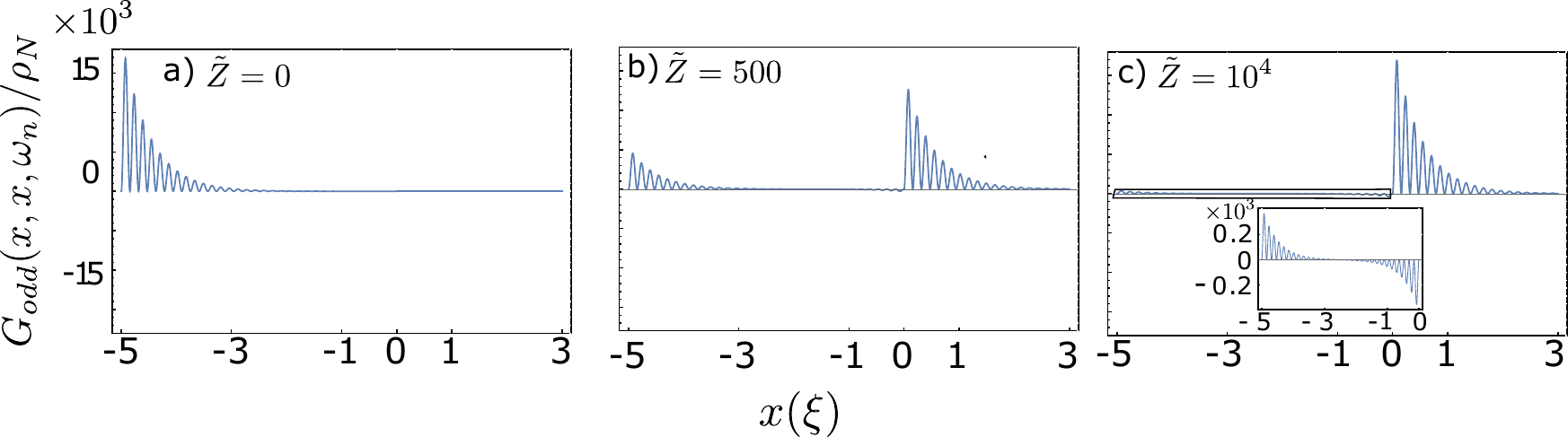}
\caption{ \label{oddfreqshort} Normalized odd-frequency component of anomalous Green's function for $L = 5 \xi$ with several values of $\tilde{Z}$ a) $\tilde{Z}=0$, b) $\tilde{Z}=500$, and c) $\tilde{Z}=10^4$. The other parameters are $\Delta_0 = 0.1\mu$ and $\omega_n =\Delta_0/1000$.
Note that the $y$-axis is given in units of $10^{3}$ (semi-infinite system with impurity).  }
\end{figure*}

After discussing the zero-energy LDOS, we can now focus on the superconducting pair correlations of the system. The bulk of a $p$-wave superconductor only consists of an even-frequency component with no odd-frequency one. However, in the presence of spatial non-uniformity, as in the present system, the odd-frequency component can be enhanced \cite{oddfreqrev1}.
Using Eq. (\ref{oddfreqeq1}), we can plot the odd-frequency $s$-wave component of the Green's function. We have analytically continued the function to the Matsubara frequency using the substitution $E +  i \delta \rightarrow i \omega_n $. This makes the odd-frequency component an odd function in frequency $\omega_n$. In Fig. \ref{oddfreqlarge} we have shown the odd-frequency Green's function for $x=x'$ as a function of $x$ for different values of impurity strengths $\Tilde{Z} $ with $L = 10 \xi$. We used $\Delta_0 = 0.1\mu$. The $x$-axis is given in units of $\xi$. In Fig. \ref{oddfreqshort} we used $L = 5 \xi$. The corresponding LDOSs are shown in Figs. \ref{ldoslarge} and \ref{ldossmall}, respectively. 

In Fig. \ref{oddfreqlarge}, as we increase the strength of the impurity potential we notice that the odd-frequency component at the impurity increases. Notably, increasing the impurity potential further did not alter the graph. Figure \ref{oddfreqlarge}(c) is similar locally around the impurity in Fig. \ref{LDOSoddfreqhomogeneous}(b). Figure \ref{oddfreqlarge} shows that for large $L$, increasing the value of the impurity strength does not affect the odd pair correlations at the edge and only enhances those at the impurity site. Therefore, the impurity does not affect the edge as there is no interference.

The interference effects can be seen in Fig. \ref{oddfreqshort}, which is plotted for a moderate value of $L$. As we increase $\tilde{Z}$, the odd-frequency component is no longer symmetric around the impurity. Upon further increasing the strength, the value of the odd-frequency component near the edge and the left-hand side of the impurity is significantly altered. The reason for this drastic change is that the sign of the odd frequency component at the edge and the impurity are opposite to each other. Provided $k_f L$ is far from $n\pi$, as one decreases the distance between the impurity and the edge, they destructively interfere to give the corresponding outcome presented in Fig. \ref{oddfreqshort}.

To further analyze Figs. \ref{oddfreqlarge} and  \ref{oddfreqshort} we can try to find the expression for the odd-frequency component in the low-energy limit. This will enable us to extract information about the localization length.
\subsection{Zero Energy Correlation functions}\label{localisationsec}
The impact of the impurity on the zero-energy state can be determined by examining the low-energy behavior of the odd-frequency component and LDOS. 
However, if we take the limit of $E \rightarrow 0$, due to the presence of the zero-energy state, the LDOS and the odd-frequency component diverge. Thus, we only consider the terms that diverge in this limit as they contribute the most at zero energy. 
We obtain the low-energy odd-frequency Green's function (valid for finite impurity strength) as follows:
\begin{widetext}
\begin{equation}\label{fulleq}
    G_{odd}(x,x,\omega_n) = \frac{2 m }{\hbar^2 k_f} \frac{\Delta_0}{\omega_n} e^{-2 x/\xi} \frac{[\sin(k_f(x+L)) + \Theta(x)\Tilde{Z}\sin(k_f x)\sin(k_f L)]^2}{e^{2L/\xi}+ \Tilde{Z}^2\sin^2(k_f L) + \tilde{Z}\sin(2k_f L)} + \mathcal{O}(\omega_{n}) . 
\end{equation}
\end{widetext}
Details of the derivation of the above-stated expression can be found in Appendix \ref{local}.
It is important to note that the numerical results in Figs. \ref{oddfreqlarge} and \ref{oddfreqshort} are evaluated for the finite Matsubara frequency and are not evaluated at the sufficiently small frequency; further, they include contributions of orders of $\omega_n$ beyond $1/\omega_n$, including $\omega_n$, $\omega_{n}^3$ \ldots and so on. We confirmed that Eq. (\ref{fulleq}) can be reproduced numerically for a much smaller value of $\omega_n$.
In the regime $e^{L/\xi}/|\sin(k_f L)| \gg \tilde{Z}$, we recover the odd-frequency component for the semi-infinite $p$-wave superconductor system (with the edge at $x=-L$). When $\tilde{Z}\gg e^{L/\xi}/|\sin(k_f L)|$, we obtain a semi-infinite $p$-wave superconductor system with the edge at $x=0$. In the intermediate regime, one finds finite odd-frequency pairings at the edge and right-hand side of the impurity. 
In all the regimes, from the expression of $G_{odd}$ in Eq. (\ref{fulleq}), we can see that the change in $\tilde{Z}$ does not alter the exponential term $e^{-2 x/\xi}$. Thus, the impurity does not alter the characteristic length scale i.e.\ the localization length of the odd-frequency component; however, it leads to the presence of zero-energy odd-frequency pairings beyond the edge (to the right-hand side of the impurity). 
In a manner similar to that given in Appendix \ref{local}, we can determine the zero-energy LDOS\@ as follows: 
\begin{widetext}
\begin{equation}\label{ldosanalytic}
    \rho(x,0+i\delta) = \frac{2 m }{\hbar^2 k_f \pi} \frac{\Delta_0}{\delta} e^{-2 x/\xi} \frac{[\sin(k_f(x+L)) + \Theta(x)\Tilde{Z}\sin(k_f x)\sin(k_f L)]^2}{e^{2L/\xi}+ \Tilde{Z}^2\sin^2(k_f L) + \tilde{Z}\sin(2k_f L)} + \mathcal{O}(\delta) . 
\end{equation}
\end{widetext}
Equation (\ref{ldosanalytic}) reproduces Figs. \ref{ldoslarge} and \ref{ldossmall}. It is also similar to the probability density obtained from the zero-energy state wave-function in Eq. (\ref{probzeroenergy}). According to the quasi-classical theory \cite{oddfreqZESLDOS1}, a finite zero-energy LDOS is a manifestation of odd-frequency pairings \cite{oddfreqZESLDOS3}. This can be seen in the plots for the LDOS given in Figs.\ref{ldoslarge} and \ref{ldossmall}, as well as their corresponding Figs. \ref{oddfreqlarge} and \ref{oddfreqshort}.
\section{Conclusion and Discussion} \label{conclusion}
After discussing the model of the semi-infinite $p$-wave superconductor with an impurity near the edge and scattering approach, we reviewed the $p$-wave superconductor of semi-infinite geometry and infinite geometry with an impurity. We showed that the bound state energy $E_b$ for an infinite system with an impurity can be given by a simple expression, Eq. (\ref{boundstateeq1}). The analytic expression for these bound states for $d$-wave superconductor junctions had been previously predicted \cite{dwave1,dwave2}, and the analytic expression for the bound states for the $p$-wave superconductor also matched the prediction. 

Using the scattering approach, we obtained the LDOS and used the analytic expression for the zero-energy state (ZES) wave function to gain a better understanding of the impact of the impurity. The position of the ZES was seen to shift from the edge to the impurity site on increasing the impurity strength $\tilde{Z}$, thereby suggesting that the ZES in the $p$-wave system was robust against impurities for high impurity strength values. Previous studies on topological systems, such as the quantum Hall system, were valid for small values of impurity strength \cite{elgart,koma} or showed that the spectrum was significantly altered for high impurity strengths \cite{Aizenman_1998}. Thus, our result adds to the current understanding of the effect of impurities in topological systems.

We showed the odd-frequency component of the anomalous Green's function for a small value of the Matsubara frequency as a function of position. We observed some enhancements near the edge and impurity sites. Decreasing the distance between the impurity and the edge resulted in interference that significantly altered the spatial dependence of the odd-frequency component of the anomalous Green's function. However, this interference did not affect the localization length of the odd-frequency component, and we showed that it was independent of the strength of the impurity potential. The observed odd-frequency spatial dependence may be experimentally measured by probing the local Josephson coupling via scanning tunneling microscopy with a superconducting tip in semiconductor nanowire systems or other proposed methods \cite{Oddfreqexp,majoranatip,Cayao2}.
Lastly, we found the analytic expression for the zero-energy correlation functions. We discovered that the expression for the odd-frequency pairing and the LDOS had the same spatial dependence \cite{spatialprofile1}, and the LDOS was qualitatively similar to the probability density obtained from the ZES wave function. 

Lately, systems, such as one-dimensional semiconductor nanowire systems, in proximity to a conventional $s$-wave superconductor in the presence of a strong magnetic field have been discussed \cite{YuvalOreg,stanescu_tewari_2013,Lutchynpaper}. The model discussed herein could be realized with strongly charged impurities or gate voltage in such a nanowire system. Thus, tuning the gate voltage can allow us to shift the position of the ZES\@.

The current method used to calculate the Green's function is performed using the quasi-classical approximation and lacks the precision needed to probe the critical behavior of the system around the quantum critical point. Numerical methods, such as those reported in other works \cite{takagi_tamura_tanaka_2020}, are suitable for tackling this problem.

\acknowledgments

We thank J. Cayao for valuable discussions. 
Y.T. and S. T. acknowledge the support from Grant-in-Aid for Scientific Research B (KAKENHI Grant No. JP18H01176). 
Y.T. is also supported by Grant-in-Aid for Scientific Research A
(KAKENHI Grant No. JP20H00131) and the JSPS Core-to-Core program Oxide Superspin International Network. A.K. acknowledges the support from  MEXT (JP) JSPJ (Grants Nos. 15K05166 and 19H01846) from the Ministry of Education, Culture, Sports, Science, and Technology, Japan. S.M. would like to thank the Sato yo International Scholarship Foundation (SISF).

\appendix

\section{Hamiltonian for \textit{p}-wave system} \label{Ham}
The outline of the procedure to derive the quasi-classical Hamiltonian is similar to that used for $d$-wave superconductivity \cite{Kashiwaya_2000,quasiclassicalbdg,tanaka2018quasiclassical}. Using Eq. (\ref{nonlocalham}) (without impurity potential),  Eq. (\ref{BDgeq})  and (\ref{wavefnnonlocal})
we obtain the following two equations:
\begin{multline} \label{orig1}
    \bigg(-\frac{\hbar^2}{2m}\frac{d^2}{dx^2}-\mu\bigg)u(x) + \int dx' \Delta(x,x') v(x') \\ = E u(x) ,
\end{multline}
\begin{multline}\label{orig2}
    \bigg(\frac{\hbar^2}{2m}\frac{d^2}{dx^2}+\mu\bigg)v(x) - \int dx' \Delta^*(x,x') u(x') \\= E v(x) . 
\end{multline}
We make a change of variables to centre of mass coordinates as $r = x - x'$, $R = (x + x')/2$ and define $\tilde{\Delta}(r,R) \equiv \Delta(x,x')$. The Fourier transform of $\tilde{\Delta}(r,R)$ is given by the following:
\begin{equation}
    \tilde{\Delta}(k,R) = \int dr \;e^{-i k r} \tilde{\Delta}(r,R).
\end{equation}
In the quasi-classical approach one defines envelope functions $\bar{u}(\hat{k},x)$ and $\bar{v}(\hat{k},x)$ by separating the rapid fluctuations of the kinetic energy term from the wave-function. One also assumes that the Cooper pair is formed on the Fermi surface. We define the envelope functions as: 
\begin{equation}\label{quasi-classical_transform}
    \begin{pmatrix}
     u(x) \\
     v(x)
    \end{pmatrix}
    \equiv  e^{i k_f \hat{k} x}\begin{pmatrix}
     \bar{u}(\hat{k},x) \\
     \bar{v}(\hat{k},x)
    \end{pmatrix}.
\end{equation}
Here, $\hat{k} = k/k_f$ and $k$ is wave number of the quasi-particle and $|k| = k_f$. 
Using Eq. (\ref{quasi-classical_transform}), the BdG equation can be rewritten as:
\begin{multline} \label{bdg1}
    \bigg(\frac{\hbar^2 k_{f}^2 }{2m} - i \hbar v_f \hat{k}\frac{d}{d x} +\frac{\hbar^2 }{2m}\frac{d^2}{d x^2} -\mu \bigg)\bar{u}(x) \\+ \int dx' \Delta(x,x') \bar{v}(x')e^{-i k(x-x')} = E \bar{u}(x) ,
\end{multline}
\begin{multline}\label{bdg2}
    \bigg(-\frac{\hbar^2 k_{f}^2}{2m} + i \hbar v_f \hat{k} \frac{d}{dx} -\frac{\hbar^2 }{2m}\frac{d^2}{dx^2}  +\mu\bigg)\bar{v}(x) \\+ \int dx' \Delta^*(x,x') \bar{u}(x')e^{-i k(x-x')} = E \bar{v}(x) . 
\end{multline}

The integral part of Eq. (\ref{bdg1}) can be rewritten as:
\begin{multline} \label{qc1}
    \int dx' \Delta(x,x') \bar{v}(x')e^{-i k(x-x')} \\= \int dr \tilde{\Delta}(r,x-r/2) \bar{v}(x-r)e^{-i k r} \\ \approx \tilde{\Delta}(k,x)\bar{v}(x).
\end{multline}
Where one obtains the last approximation after Taylor expansion up to the zeroth order \cite{quasiclassicalbdg}.
Similarly, one can write the expression for the integral in Eq. (\ref{bdg2}) as:
\begin{multline}\label{qc2}
    \int dx' \Delta^*(x,x') \bar{u}(x')e^{-i k(x-x')}  \approx \tilde{\Delta}^*(-k,x)\bar{u}(x).
\end{multline}
 Relabelling $\tilde{\Delta}(k,x)$ as $\Delta(\hat{k},x)$, dropping the second derivative terms and using Eqs. (\ref{qc1}) and (\ref{qc2}), one can rewrite Eq. (\ref{orig1}) and Eq. (\ref{orig2}) as follows: 
\begin{equation}
    - i\hbar v_f \hat{k} \frac{d}{dx} \bar{u}(\hat{k}, x) + \Delta(\hat{k},x) \bar{v}(\hat{k}, x) = E \bar{u}(\hat{k}, x),
\end{equation}
\begin{equation}
     i\hbar v_f \hat{k} \frac{d}{dx} \bar{v}(\hat{k}, x) - \Delta^*(-\hat{k},x) \bar{u}(\hat{k}, x) = E \bar{v}(\hat{k}, x)
\end{equation}

The phase factor of $\Delta$ does not play important role here and we set $\Delta$ as a real function of $\hat{k}$ and $x$. We use the relation $-\Delta^*(-\hat{k},x) =\Delta(\hat{k},x)$, which is valid  for \textit{p}-wave spin-triplet superconductors \cite{pwaveandreev} for a real pair potential.
Then, we can write a single particle quasi-classical Hamiltonian as: 

\begin{equation} \label{QChamil}
    \hat{H}_{QC}(\hat{k},x) = -\bigg(i \hbar v_f \hat{k} \frac{d}{d x} \bigg)\sigma_z + \Delta(\hat{k},x)\sigma_x  . 
\end{equation}

For $p$-wave superconductivity we choose\cite{PhysRevB.99.184512},
\begin{equation} \label{pairpotqc}
    \Delta(\hat{k},x) = 
   \Delta_0 \frac{k}{\sqrt{k^2}} . 
\end{equation}
We use this pair potential for calculations throughout this paper.

\section{Wave functions}\label{wavefunctions}
\begin{figure}[b]
\includegraphics[width = 8.5 cm]{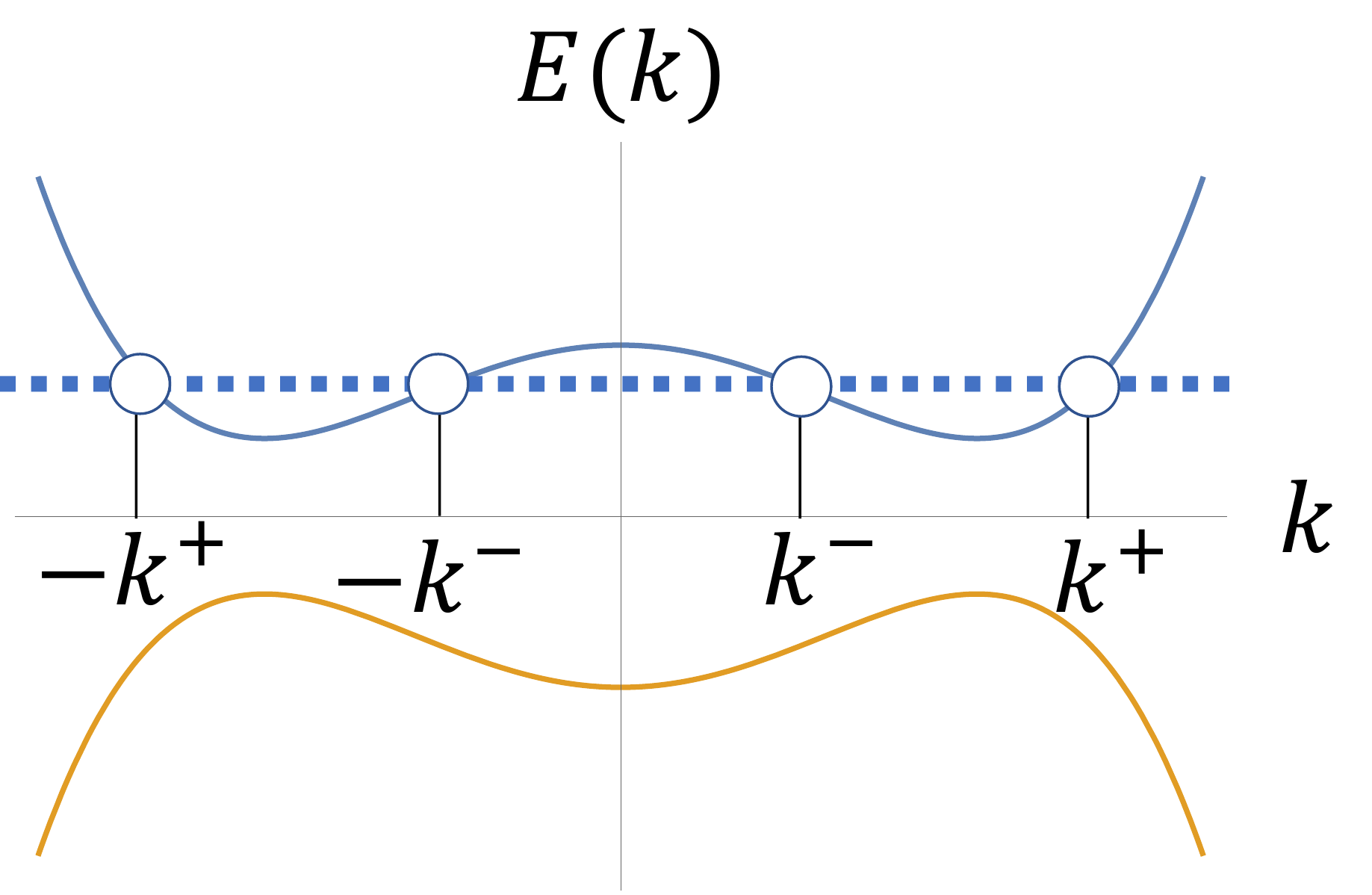}
\caption{\label{disp} Schematic picture of dispersion relation of superconductor.}
\end{figure}

For a spin-less $p$-wave superconductor the pair potential, given by Eq. (\ref{pairpotqc}), depends on the direction of the wave vector \textit{k}.
There are four possible wave functions for a given energy $E$ as indicated in the dispersion relation in Fig. \ref{disp}.
Solving the eigenvalue equation for $\hat{H}_{QC}$ and then using Eq. (\ref{quasi-classical_transform}), we get the following wave-functions of $\hat{H}(x,x')$ in the coordinate basis:
\begin{eqnarray}
    \Psi_{ k^+}(x) &=& e^{ ik^+x}
    \begin{pmatrix}
    1 \\ \Gamma
    \end{pmatrix} ,  \\
    \Psi_{ -k^+}(x) &=& e^{ -ik^+x}
    \begin{pmatrix}
    1 \\ -\Gamma
    \end{pmatrix} , \\
    \Psi_{ k^-}(x) &=& e^{ ik^-x}
    \begin{pmatrix}
    \Gamma \\ 1
    \end{pmatrix} , \\
    \Psi_{ -k^-}(x) &=& e^{ -ik^-x}
    \begin{pmatrix}
    -\Gamma \\ 1
    \end{pmatrix} . 
\end{eqnarray}
Here, within the quasi-classical approximation $k^\pm$ is given by, 
\begin{equation}
    k^\pm \approx k_f \pm \gamma(E) . 
\end{equation} 
with $\gamma(E) = \frac{k_f \Omega(E)}{2\mu}$, $\Omega(E) = \sqrt{E^2 - \Delta_0^2}$ and $\Gamma(E)= \frac{\Delta_0}{E + \Omega(E)}$. 
Similarly, The eigenvectors of the transpose of $\hat{H}(x,x')$ within the quasi-classical approximation are as follows:   
\begin{eqnarray}
    \Tilde{\Psi}_{ k^+}(x) = e^{ ik^+x}
    \begin{pmatrix}
    1 \\ -\Gamma
    \end{pmatrix}  , \\
    \Tilde{\Psi}_{ -k^+}(x) = e^{ -ik^+x}
    \begin{pmatrix}
    1 \\\Gamma
    \end{pmatrix} , \\
    \Tilde{\Psi}_{ k^-}(x) = e^{ ik^-x}
    \begin{pmatrix}
    -\Gamma \\ 1
    \end{pmatrix} , \\
    \Tilde{\Psi}_{ -k^-}(x) = e^{ -ik^-x}
    \begin{pmatrix}
    \Gamma \\ 1
    \end{pmatrix} . 
\end{eqnarray}

We can now define the scattering states of the system using the  found wave functions. 

\section{Scattering States} \label{scatteringstates}

We can now define scattering states for this system similar to \cite{FURUSAKI1991299}. We have four possible scattering states as given below:
\begin{widetext}
\begin{eqnarray}
    \Psi_{out}^{(+)}(x) &=&
        \begin{cases}
         \Psi_{k^+}(x)+ a_1\Psi_{k^-}(x)+b_1\Psi_{-k^+}(x),& -L<x<0\\
         c_1\Psi_{k^+}(x) + d_1\Psi_{-k^-}(x),& x>0
        \end{cases}\\
        \Psi_{out}^{(-)}(x)&=&
        \begin{cases}
         \Psi_{-k^-}(x)+ a_2\Psi_{-k^+}(x)+b_2\Psi_{k^-}(x),& -L<x<0\\
         c_2\Psi_{-k^-}(x) + d_2\Psi_{k^+}(x),& x>0
        \end{cases}\\
        \Psi_{in}^{(+)}(x)&=&
        \begin{cases}
         c_3\Psi_{-k^+}(x)+ d_3\Psi_{k^-}(x)+e_3\Psi_{k^+}(x)+ f_3 \Psi_{-k^-}(x),& -L<x<0\\
         \Psi_{-k^+}(x) + a_3\Psi_{-k^-}(x)+ b_3\Psi_{k^+}(x),& x>0
        \end{cases}\\
        \Psi_{in}^{(-)}(x)&=&
        \begin{cases}
         c_4\Psi_{k^-}(x)+ d_4\Psi_{-k^+}(x)+e_4\Psi_{-k^-}(x)+ f_4 \Psi_{k^+}(x),& -L<x<0\\
         \Psi_{k^-}(x) + a_4\Psi_{k^+}(x)+ b_4\Psi_{-k^-}(x).& x>0
        \end{cases}
\end{eqnarray}
The figures for the scattering states for $\Psi_{out(in)}^{(\pm)}(x)$ has been drawn in Figs. \ref{scatteringstatesfig} (a)-(d), respectively.
\begin{figure}
    \centering
    \includegraphics[width = 14 cm]{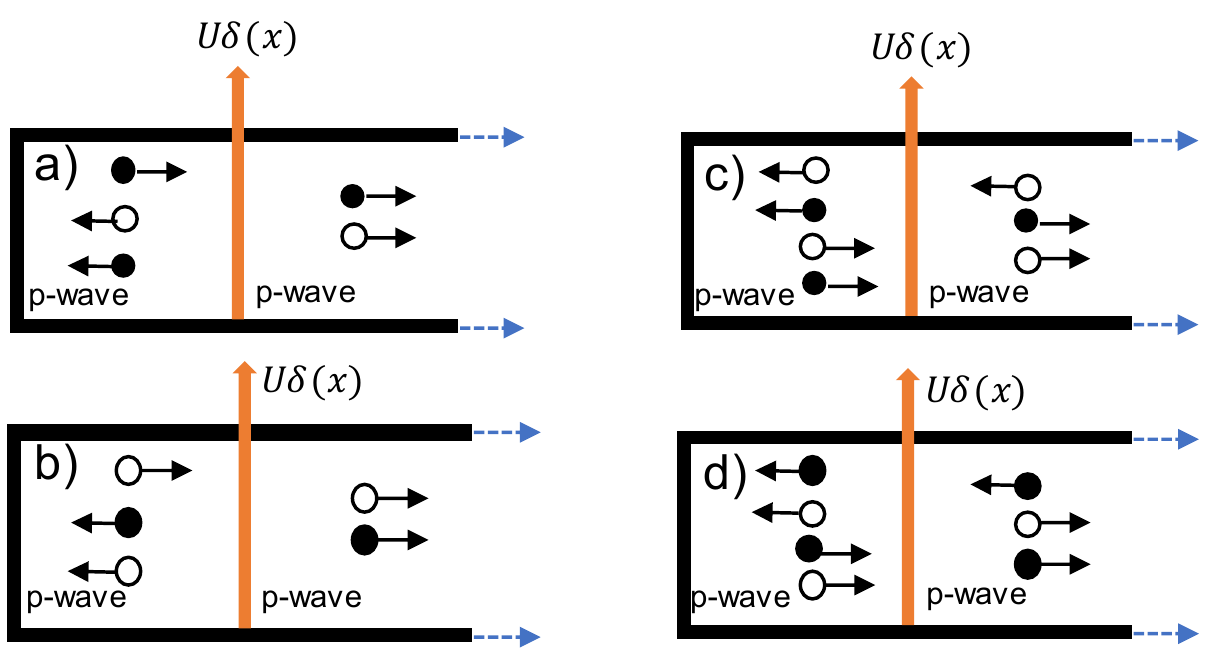}
    \caption{Schematic illustration of the scattering processes where filled (empty) circle indicates electron like quasi-particle (hole like quasi-particle).}
    \label{scatteringstatesfig}
\end{figure}
\end{widetext}
In the above equations
$a_i$ and $b_i$ for $i=1,\ldots,4$ represent the Andreev and normal reflection coefficients respectively. $c_i$ and $d_i$ for $i=1,\ldots,4$ represent the transmission coefficients through the delta function impurity at $x=0$. $f_i$ and $e_i$ for $i=3,4$ are the reflection coefficients for the waves scattered from the edge at $x=-L$.
We can also define the scattering states for the conjugate processes that are eigenstates of the Hamiltonian $H^t$ where the superscript denotes the transpose of the Hamiltonian. The states are as follows:
\begin{widetext}
\begin{eqnarray}
    \Tilde{\Psi}_{out}^{(+)}(x) &=&
        \begin{cases}
         \Tilde{\Psi}_{k^+}(x)+ \Tilde{a}_1\Tilde{\Psi}_{k^-}(x)+\Tilde{b}_1\Tilde{\Psi}_{-k^+}(x),& -L<x<0\\
         \Tilde{c}_1\Tilde{\Psi}_{k^+}(x) + \Tilde{d}_1\Tilde{\Psi}_{-k^-}(x),& x>0
        \end{cases} \\
         \Tilde{\Psi}_{out}^{(-)}(x)&=&
        \begin{cases}
         \Tilde{\Psi}_{-k^-}(x)+ \Tilde{a}_2\Tilde{\Psi}_{-k^+}(x)+\Tilde{b}_2\Tilde{\Psi}_{k^-}(x),& -L<x<0\\
         \Tilde{c}_2\Tilde{\Psi}_{-k^-}(x) + \Tilde{d}_2\Tilde{\Psi}_{k^+}(x),& x>0
        \end{cases}\\
        \Tilde{\Psi}_{in}^{(+)}(x)&=&
        \begin{cases}
         \Tilde{c}_3\Tilde{\Psi}_{-k^+}(x)+ \Tilde{d}_3\Tilde{\Psi}_{k^-}(x)+\Tilde{e}_3\Tilde{\Psi}_{k^+}(x)+ \Tilde{f}_3 \Tilde{\Psi}_{-k^-}(x),& -L<x<0\\
         \Tilde{\Psi}_{-k^+}(x) + \Tilde{a}_3\Tilde{\Psi}_{-k^-}(x)+ \Tilde{b}_3\Tilde{\Psi}_{k^+}(x),& x>0
        \end{cases}\\
         \Tilde{\Psi}_{in}^{(-)}(x)&=&
        \begin{cases}
         \Tilde{c}_4\Tilde{\Psi}_{k^-}(x)+ \Tilde{d}_4\Tilde{\Psi}_{-k^+}(x)+\Tilde{e}_4\Tilde{\Psi}_{-k^-}(x)+ \Tilde{f}_4 \Tilde{\Psi}_{k^+}(x),& -L<x<0\\
         \Tilde{\Psi}_{k^-}(x) + \Tilde{a}_4\Tilde{\Psi}_{k^+}(x)+ \Tilde{b}_4\Tilde{\Psi}_{-k^-}(x).& x>0
        \end{cases}
\end{eqnarray}
\end{widetext}
In the above equations
$\tilde{a}_i$ and $\tilde{b}_i$ for $i=1,\ldots,4$ represent the Andreev and normal reflection coefficients respectively. $\tilde{c}_i$ and $\tilde{d}_i$ for $i=1,\ldots,4$ represent the transmission coefficients through the delta function impurity at $x=0$. $\tilde{f}_i$ and $\tilde{e}_i$ for $i=3,4$ are the reflection coefficients for the waves scattered from the edge at $x=-L$. Coefficients can be found by imposing continuity of the wave function for both incoming and outgoing scattering states at $x=0$ along with derivative condition for delta function potential $U\delta(x)$ as follows:
\begin{eqnarray}
    \frac{d}{dx}\Psi_{out(in)}^{\pm}(x)\bigg|_{0^-}^{0+} = \frac{2m}{\hbar^2}U \Psi_{out(in)}^{\pm}(0) , \\
    \frac{d}{dx}\Tilde{\Psi}_{out(in)}^{\pm}(x)\bigg|_{0^-}^{0+} = \frac{2m}{\hbar^2}U \tilde{\Psi}_{out(in)}^{\pm}(0) . 
\end{eqnarray}
Here, $U$ is the strength of the delta function impurity at $x=0$. 

Incoming scattering states follow incoming boundary conditions i.e.\ $\Psi^{(\pm)}_{in}(-L)=0$ and $\tilde{\Psi}^{(\pm)}_{in}(-L)=0$ and outgoing scattering states follow the outgoing boundary condition that at $+\infty$ the wave is asymptotic to a plane wave.

\section{Scattering coefficients}\label{scatteringcoef}
\begin{widetext}
Using the continuity and differentiability conditions at $x=0$ along with the incoming boundary condition $\Psi^{(\pm)}_{in}(-L)=0$, we obtain the following values of the coefficients within the quasi-classical approximation (let $Z = \frac{2m}{\hslash^2}U$):
Let $D_1$ be,
\begin{equation}
    D_1 = 4k_f^2(-1+\Gamma^2)^2 + Z^2(1+ \Gamma^2)^2 . 
\end{equation}
Then,
\begin{eqnarray}
a_1 = -a_2 &=& -\frac{2Z^2}{D_1}\Gamma (1+\Gamma^2) , \\
b_1 &=& \frac{Z(2 i k_f + Z)}{D_1}(-1+\Gamma^4) , \\
c_1 &=&- \frac{ 2 i k_f(2 i k_f + Z)}{D_1} (-1+\Gamma^2)^2 , \\
d_1 =d_2&=& -\frac{4 i k_f Z}{D_1} \Gamma(-1+\Gamma^2)^2 , \\
b_2 &=& -\frac{Z(2 i k_f - Z)}{D_1} (-1+\Gamma^4) , \\
c_2 &=& -\frac{2 i k_f (2 i k_f - Z)}{D_1}(-1+\Gamma^2)^2 . 
\end{eqnarray}
Let $D_2$ be,
\begin{multline}
    D_2 = -8 e^{
     2 i k_f L}Z^2 \Gamma^2 + 
   i Z(2 k_f + i Z) (-1 + \Gamma^2)^2 - 
   e^{4 i k_f L} Z (2 i k_f + Z) (-1 + \Gamma^2)^2 \\+ 
   e^{2 i L (k_f + \gamma)}Z^2 (1 + \Gamma^2)^2 + 
   e^{2 i L (k_f - \gamma)} D_1 . 
\end{multline}
Then,
\begin{eqnarray}
c_3 &=& \frac{2k_f[e^{2 i L (k_f - \gamma)}(2 k_f - i Z) + 
     i Z]}{D_2}  (-1 + \Gamma^2)^2 , \\
d_3 = d_4 &=&  \frac{4 i k_f Z e^{i L (k_f - \gamma)}}{D_2} [e^{i L (k_f - \gamma)} - 
    e^{i L (k_f + \gamma)}]  \Gamma (-1 + \Gamma^2)   ,  \\
e_3&=&-\frac{4 i k_f[e^{2 i k_f L} (2 i k_f + Z) - Z] }{D_2 (1 + \Gamma^2)} \Gamma (-1 + \Gamma^2)^2 ,   \\
f_3 &=& -\frac{2 i e^{2 i k_f L} k_f}{D_2 (1 + \Gamma^2)} [4 Z \Gamma^2 + 
    e^{2 i k_f L} (2 i k_f + Z) (-1 + \Gamma^2)^2 -
    e^{2 i L \gamma} Z (1 + \Gamma^2)^2] (-1 + \Gamma^2) ,
\end{eqnarray}
\begin{equation}
a_3 = - a_4  = \bigg( \frac{2\Gamma}{1 + \Gamma^2}\bigg) + \bigg( \frac{2\Gamma}{1 + \Gamma^2}\bigg)             \frac{4 k_f^2 e^{2 i L k_f}}{D_2 }[e^{2 i \gamma L}+e^{-2 i \gamma L}]\Gamma(-1 + \Gamma^2)^2 ,
\end{equation}
\begin{multline}
b_3 = \bigg(\frac{-1 + \Gamma^2}{1+ \Gamma^2}\bigg)+ \bigg(\frac{-1 + \Gamma^2}{1+ \Gamma^2}\bigg)\frac{1}{D_2} \{-16 i k_f Z e^{2 i k_f L} \Gamma^2 + Z^2 e^{2 i L(k_f + \gamma) }(1+ \Gamma^2)^2 \\- e^{2 i L (k_f - \gamma)}[4 k_{f}^2 (-1 + \Gamma^2)^2 -2 i k_f Z (1 + \Gamma^2)^2 ]\},
\end{multline}

\begin{eqnarray}
c_4 &=& \frac{2 e^{2 i L (k_f - \gamma)} k_f}{D_2} [2 k_f - 
    i (-1 + e^{2 i L (k_f + \gamma)}) Z] (-1 + \Gamma^2)^2 ,  \\
e_4 &=& \frac{4 i k_f e^{2 i k_f L} [2 i k_f + (-1 + e^{2 i k_f L}) Z] }{D_2 (1 + \Gamma^2)}\Gamma (-1 + \Gamma^2)^2 , \\
f_4 &=& \frac{2 k_f \{2 k_f (-1 + \Gamma^2)^2 - 
    i Z [-4 e^{2 i k_f L} \Gamma^2 - (-1 + \Gamma^2)^2 + 
       e^{2 i L (k_f + \gamma)} (1 + \Gamma^2)^2]\}}{D_2 (1 + \Gamma^2)} (-1 + \Gamma^2) ,   
\end{eqnarray}
\begin{multline}
    b_4 =\bigg(\frac{-1 + \Gamma^2}{1+ \Gamma^2}\bigg)+ \bigg(\frac{-1 + \Gamma^2}{1+ \Gamma^2}\bigg)\frac{1}{D_2} \{(4 k_{f}^2 + Z^2)(-1 + \Gamma^2)^2  + 16 i k_f Z \Gamma^2 e^{2 i k_f L} \\ + 2 i k_f Z (-1 + \Gamma^2)^2 e^{4 i k_f L} - 2 i k_f Z e^{2 i L (k_f +\gamma)}(1 + \Gamma^2)^2
    -e^{2 i L (k_f -\gamma)}[4 k_{f}^2 (-1 + \Gamma^2)^2 + 2 i k_f Z (1 + \Gamma^2 )^2]\}. 
\end{multline}

$D_1 = 0 $ and $D_2 =  0 $ represent singular points which occur at specific energies and impurity strengths. For $D_1 = 0 $ we obtain Eqs. (\ref{boundstateeq2}) and (\ref{boundstateeq1}):
\begin{equation*}
    E_b(\Tilde{Z}) = \Delta_0 \sqrt{\sigma_N(\Tilde{Z}}) ,
\end{equation*}
with, 
\begin{equation*}
    \sigma_N(\Tilde{Z}) = \frac{4}{4 + \Tilde{Z}^2} . 
\end{equation*}
Where $\Tilde{Z} = Z/k_f$ and $E_b$ represent the bound state energy.
For the condition $D_2 = 0 $ a simple expression cannot be found but for the limit of large $Z$, one obtains the following relation:
\begin{equation}\label{quantisationeq}
E^2 [\cos (2 k_f L)-\cos (2 \gamma L)]+2 \Delta_0 ^2 \sin ^2(2 k_f L) = 0 . 
\end{equation}
Note, $\gamma$ is a function of energy. Equation (\ref{quantisationeq}) represents the quantization of energy above the superconducting gap $\Delta_0$ for large values of impurity strength $\tilde{Z} $($\tilde{Z}\gg 1$).

The scattering coefficients for the conjugate processes can be found with a similar procedure. In the quasi-classical approximation, the following relations hold:
\begin{equation}
    \Tilde{a}_1 = -a_1, \;  \Tilde{a}_2 = a_1,\; \Tilde{a}_3 = -a_3,\;     \Tilde{a}_4 = a_3 , 
\end{equation}

\begin{equation}
    \Tilde{b}_1 = b_1,\;\Tilde{b}_2 = b_2, \;  \Tilde{b}_3 = b_3,\; \Tilde{b}_4 = b_4\; , 
\end{equation}
\begin{equation}
    \Tilde{c}_1 = c_1, \; \Tilde{c}_2 = c_2, \; \Tilde{c}_3 = c_3, \; \Tilde{c}_4 = c_4\; , 
\end{equation}
\begin{equation}
    \Tilde{d}_1 = -d_1, \; \Tilde{d}_2 = -d_2, \; \Tilde{d}_3 = -d_3, \; \Tilde{d}_4 = -d_4 , 
\end{equation}
\begin{equation}
    \Tilde{e}_3 = -e_3, \;\Tilde{e}_4 = -e_4, \; \Tilde{f}_3 = f_3, \; \Tilde{f}_4 = f_4 . 
\end{equation}

Note that $d_3,\Tilde{d}_3,d_4$ and $\Tilde{d}_4$ are small and are approximated as 0 while finding the coefficients of the Green's function.
The above relations are used to find the coefficients of the Green's function.
\end{widetext}
\section{Green's function} \label{greensfn}
We can write the retarded Green's function \cite{FURUSAKI1991299} as the following:
\begin{widetext}
\begin{equation}
    G^r(x,x',E) = \\
    \begin{cases}
    \alpha_1\Psi_{out}^{(+)}(x)\Tilde{\Psi}_{in}^{(+) t}(x')+\alpha_2\Psi_{out}^{(+)}(x)\Tilde{\Psi}_{in}^{(-) t}(x')+\alpha_3\Psi_{out}^{(-)}(x)\Tilde{\Psi}_{in}^{(+) t}(x')+\alpha_4\Psi_{out}^{(-)}(x)\Tilde{\Psi}_{in}^{(-) t}(x') ,& x>x' \\
    \beta_1\Psi_{in}^{(+)}(x)\Tilde{\Psi}_{out}^{(+) t}(x')+\beta_2\Psi_{in}^{(+)}(x)\Tilde{\Psi}_{out}^{(-) t}(x')+\beta_3\Psi_{in}^{(-)}(x)\Tilde{\Psi}_{out}^{(+) t}(x')+\beta_4\Psi_{in}^{(-)}(x)\Tilde{\Psi}_{out}^{(-) t}(x').& x<x'
    \end{cases}
\end{equation}
\end{widetext}
The values of the coefficients can be obtained by using the continuity of the Green's function at $x=0$ and the derivative condition as given below: 
\begin{multline}\label{diffocnd}
    \frac{\partial}{\partial x}G^r(x,x',E)\bigg|_{x = x'^+}- \frac{\partial}{\partial x}G^r(x,x',E)\bigg|_{x=x'^-} \\= \frac{2m}{\hslash^2}\sigma_z  . 
\end{multline}
Under the quasi classical approximation\cite{PhysRevB.83.224511} we can use the relation: $k^\pm \approx k_f \pm \gamma$ with $\gamma = \frac{k_f\Omega}{2\mu}$, $k_f = \sqrt{\frac{2m}{\hslash^2}\mu}$, $\Omega(E) = \sqrt{E^2 - \Delta_0^2}$ and $\Gamma(E)= \frac{\Delta_0}{E + \Omega(E)}$.

By using the continuity of the Green's function and Eq. (\ref{diffocnd}), we obtain the following coefficients:
\begin{eqnarray}
    \alpha_2 &=& \alpha_3=-\beta_2=-\beta_3  = -\frac{2 m}{\hslash^2}\frac{ \Gamma  Z}{2 (\Gamma ^2-1)^2 {k_f}^2} , \\
    \alpha_1 &=&\beta_1 = \frac{2 m}{\hslash^2}\frac{ (-Z+2 i {k_f})}{4 (\Gamma^2-1) {k_f}^2} , \\
    \alpha_4 &=&\beta_4 = \frac{2 m}{\hslash^2}\frac{ (Z+2 i {k_f})}{4 (\Gamma^2-1) {k_f}^2} ,
\end{eqnarray}
with,
\begin{equation}
    Z = \frac{2 m}{\hslash^2}U . 
\end{equation}
Where $U$ is the strength of the delta potential at $x=0$.
Thus, we have found the Green's function for the semi-infinite $p$-wave superconductor with a nonmagnetic impurity near the edge modelled by a $\delta$ function.
We can obtain the advanced Green's function from the retarded Green's function using the relation as follows,
\begin{equation} \label{oddfreqeq2}
    G^{a}(x,x',E) = (G^r(x',x,E))^\dagger . 
\end{equation}
If we obtain the advanced Green's function 
$\bm{G}^{a}(x,x',E)$ and obtain 
$G_{odd(even)}^{a}(x,x',E)$ similar to $G_{odd(even)}^{r}(x,x',E)$ in Eqs. (\ref{oddfreq1}) and (\ref{oddfreq2}), 
we obtain the following relations, 
\begin{equation}
    G_{even}^{a}(x,x',-E) = G_{even}^{r}(x,x',E) , 
\end{equation}
\begin{equation}
    G_{odd}^{a}(x,x',-E) = - G_{odd}^{r}(x,x',E) . 
\end{equation}
\section{Zero energy odd-frequency pairing and localisation length} \label{local}
We shall consider the case $-L<x<0$ first. The limit $E\rightarrow0$ corresponds to $\Gamma\rightarrow-i$. It is easier to work with this limit. In this limit, $a_1$, $b_1$, $a_2$ and $b_2$ are 0. This will also simplify the equations. In the quasi-classical approximation we get,
\begin{widetext}
\begin{multline}\label{oddfreqdivided}
    G_{odd}(x,x,E) =  -i c_3 \alpha_1 - i d_3 \alpha_2 + i d_4 \alpha_2 + i c_4 \alpha_4+  e^{2 i k_f x} (-d_3 \alpha_1 + c_4 \alpha_2) + e^{-2 i k_f x} (c_3 \alpha_2 - d_4 \alpha_4) \\ + i e^{2 i x (k_f + \gamma)} (f_3 \alpha_1 - e_4 \alpha_2) +
 e^{2 i x \gamma} (-e_3 \alpha_1 - f_3 \alpha_2 + f_4 \alpha_2 + e_4 \alpha_4) - i e^{-2 i x (k_f - \gamma)} (e_3 \alpha_2 - f_4 \alpha_4) . 
\end{multline}
\end{widetext}
Since the expressions $e_3$, $e_4$, $f_3$ and $f_4$ ($c_3$, $c_4$, $d_3$ and $d_4$) diverge (do not diverge) in the limit $\Gamma \rightarrow -i$, we can separate Eq. (\ref{oddfreqdivided}) into two parts: $G_{odd}(x,x,E) = D(x,x,E)+B(x,x,E) $ where $D(x,x,E)$ diverges in the limit and $B(x,x,E)$ does not diverge in the limit. $D(x,x,E)$ and $B(x,x,E)$ are given by 
\begin{widetext}
\begin{equation}
    D(x,x,E)  = -i c_3 \alpha_1 - i d_3 \alpha2 + i d_4 \alpha_2 + i c_4 \alpha_4+ e^{2 i k_f x} (-d_3 \alpha_1 + c_4 \alpha_2)+e^{-2 i k_f x} (c_3 \alpha_2 - d_4 \alpha_4)  , 
\end{equation}
\begin{equation}
    B(x,x,E)=   i e^{2 i x (k_f + \gamma)} (f_3 \alpha_1 - e_4 \alpha_2)  +
 e^{2 i x \gamma} (-e_3 \alpha_1 - f_3 \alpha_2 + f_4 \alpha_2 + e_4 \alpha_4) - i e^{-2 i x (k_f - \gamma)} (e_3 \alpha_2 - f_4 \alpha_4) . 
\end{equation}
\end{widetext}

The diverging term survives for finite $\Tilde{Z}$ and we obtain,
\begin{widetext}
\begin{equation}\label{lefteq}
    G_{odd}(x,x,E) = \frac{2 m i}{\hbar^2 k_f} \frac{\Delta_0}{E} e^{-2 x/\xi} \frac{\sin^2(k_f(x+L)) }{e^{-2i\gamma L}+ \Tilde{Z}^2\sin^2(k_f L) + \tilde{Z}\sin(2k_f L)} . 
\end{equation}
\end{widetext}
A similar treatment can be done for the case of $x>0$ and we obtain the following equation:
\newpage 
\begin{widetext}
\begin{equation} \label{righteq}
     G_{odd}(x,x,E) = \frac{2 m i}{\hbar^2 k_f} \frac{\Delta_0}{E} e^{-2 x/\xi} \frac{[\sin(k_f(x+L)) + \Tilde{Z}\sin(k_f x)\sin(k_f L)]^2}{e^{2 L/\xi}+ \Tilde{Z}^2\sin^2(k_f L) + \tilde{Z}\sin(2k_f L)} . 
\end{equation}
\end{widetext}

\begin{widetext}
Combining Eqs. \ref{lefteq} and \ref{righteq} and analytic continuation to Matsubara frequency, we obtain a compact equation given by Eq. (\ref{fulleq}) of the main text as follows:
\begin{equation*}
    G_{odd}(x,x,\omega_n) = \frac{2 m}{\hbar^2 k_f} \frac{\Delta_0}{\omega_n} e^{-2 x/\xi} \frac{[\sin(k_f(x+L)) + \Theta(x)\Tilde{Z}\sin(k_f x)\sin(k_f L)]^2}{e^{2 L/\xi}+ \Tilde{Z}^2\sin^2(k_f L) + \tilde{Z}\sin(2k_f L)} . 
\end{equation*}

\end{widetext}
\bibliographystyle{apsrev4-2}
\bibliography{bib}
\end{document}